\documentclass[pre,a4paper,twocolumn,showpacs,amsmath]{revtex4}
\usepackage{graphicx,amsfonts,amsmath}

\newcommand{\sgn}{\mathop{\rm sgn}\nolimits}

\begin{document}

\title{Quasi-rigidity: some uniqueness issues}
\author{Sean C. McNamara}
\affiliation{Institut f\"ur Computerphysik, Universit\"at Stuttgart,
70569 Stuttgart, GERMANY}

\author{Hans J. Herrmann}
\affiliation{Institut f\"ur Baustoffe, HIF E12, ETH H\"onggerberg CH-8093,
Z\"urich, SWITZERLAND}

\date{\today}

\begin{abstract}
Quasi-rigidity means that one builds a theory for assemblies of
grains under a slowly changing external load
by using the deformation of those grains as a small parameter.
Is quasi-rigidity a complete theory for these granular assemblies?  
Does it provide unique predictions of the assembly's behavior, or 
must some other process be invoked to decide between several possibilities?
We provide evidence that quasi-rigidity is a complete theory by showing
that two possible sources of indeterminacy do not exist for the case
of disk shaped grains.  One possible source of indeterminacy arises from
zero-frequency modes present in the packing.  This problem
can be solved by considering the conditions required to obtain force
equilibrium.  A second possible source of indeterminacy is the necessity
to choose the status (sliding or non-sliding) at each contact.  We show
that only one choice is permitted, if contacts slide
only when required by Coulomb friction.
\end{abstract}

\maketitle

\section{Introduction}

\subsection{Historical Overview}

The foundation of many physical theories is the observation that a
certain physical quantity is ``small''.  In practice, this means that
the ratio between two different quantities with the same units
is much less than unity.  Once a small quantity has
been identified, there are two ways of proceeding.  First of all,
that quantity can be set to zero, if one wishes to emphasize other aspects
of the system.  This is what is done when presenting the harmonic oscillator
to students for the first time: one usually sets the dissipation to zero,
even though its effects are quite important.  The second possibility is
to use the small quantity to linearize the equations.
This is what one does when one suspects that the quantity, though small,
plays an important role.  An example is linear stability analysis.

In the study of granular
materials, an obvious choice for a small quantity is the distance particles
must move in order to activate contact forces.  This choice is motivated
by the common observation that in a pile of stones or marbles,
the deformation of the particles due to the stresses put on them is 
not visible to the naked eye.  Should these deformations be set to zero,
or kept as a small parameter? 

At the end of the last century, this
question was quite controversial, as one can see from
browsing through a conference proceeding from that time \cite{Cargese}.
The issue at hand was the stress distribution at the bottom of a
sand pile.  Several authors \cite{qmodel,Bouchaud,Claudin,Cates}
proposed theories where the grains were assumed to be perfectly
rigid.  In this way, they could circumvent the
question of a stress-free reference state.  However,
these theories were criticized on many points \cite{Savage}.
For example, it was pointed out that continuum mechanics could also account
for the observations \cite{Florence,Goddard}.  Another objection
was that they used arbitrary ways to resolve the problem of contact
force indeterminacy.  This problem arises because it is impossible
to deduce the contact forces in a static granular packing from assuming
force equilibrium.  There is no unique solution; instead many solutions
are possible \cite{Radjai96,Wolf,indet}.  This loss of uniqueness
occurs because there are more unknowns (contact forces) than equations
(vanishing force and torque on each particle).

But the root of all these objections was the realization that rigid
particle theories must be radically different from continuum mechanics.  In
continuum mechanics, the counterpart to the particle deformations
is the strain.  Thus neglecting deformations corresponds to
eliminating the strain.  But strain is a fundamental quantity,
and eliminating it destroys the entire structure of continuum mechanics.  
Continuum mechanics can describe the macroscopic behavior
of granular materials, and it would be quite strange if the best
microscopic or grain-level theory were incompatible with it.
It was no coincidence that most
opposition to rigid particle theories came from engineers, who are more
familiar with continuum mechanics than most physicists.

The recently proposed force network ensemble \cite{FNensemble} can
be considered as a modern version of the rigid particle
theories.  Instead of adding assumptions to determine the
forces, all possibilities are considered.  The system is thus
represented by a point in a high-dimensional space \cite{indet,Wolf},
much like in statistical mechanics.  One should also mention the
widely used
numerical method of contact dynamics \cite{CD}, which is also based
on the assumption of rigid particles.

This paper is concerned with an alternative approach, where the
particle deformations are a small parameter of central importance. 
We call this
approach ``quasi-rigidity''.  This choice is motivated both by the belief
that such deformations determine the contact forces in physical systems,
and by the desire to propose a theory compatible with continuum
mechanics, in view of arranging a future marriage between microscopic
and macroscopic theories.

\subsection{Quasi-rigidity}

Quasi-rigidity was first proposed as the basis
for a numerical method \cite{Kishino88,Bagi93,Goddard95,Oron,Kishino},
but has recently been explored theoretically by a number of authors
\cite{Roux,Combe,Combethese,CombePRL,Miehe,one,Kuhn,PG,Baginew}.  
Results of this approach include
a deeper understanding of isostatic packings of frictionless
particles \cite{Roux,Combethese,CombePRL}, the microscopic origins of
strain \cite{Roux,Combe,Combethese}, the stability of packings
\cite{one,Kuhn}, the relation between softening and sliding 
contacts \cite{PG}, and jamming \cite{Baginew}.
Work on a corresponding numerical method 
\cite{Combe,Combethese,Miehe} has also continued.

In quasi-rigid theories, the state of the packing is given by the
deformation of the particles at each contact. 
These deformations determine
the contact forces, which in turn govern the motion of the particles.
Finally, the particle motion gives the change of the deformations.
When studying the
response of a packing to an external load, these deformations
must be given initial values, analogous to specifying a
reference state in continuum mechanics. 

Do these theories predict a unique evolution of the packing?
Two possible sources of non-uniqueness have been pointed out \cite{Kuhn}. 
First of all, indeterminacy can occur
if there exists a possible motion that would not modify the contact
forces.  Such motions are called ``floppy modes'' or ``mechanisms''.
We call this mechanism indeterminacy.  Secondly, indeterminacy
can arise due to the necessity to choose the status (sliding or non-sliding)
of each contact.  This is contact status indeterminacy.  

In this paper, we show that both types of indeterminacy do not occur.
Mechanism instability appears to be a problem because force equilibrium
is the usual starting point for quasi-rigid theories.  But force
equilibrium can be considered as a certain limit of Newton's second law.
When this is done, one can assess the impact of any mechanisms that
may be present.  Contact status indeterminacy can be eliminated by
requiring that the contacts obey Coulomb friction, and letting
contacts slide only when necessary.
This is what is commonly done in numerical simulations.
A precise definition of `necessary' will be given later in this paper.

A third possible source of indeterminacy is opening or closing contacts.
In this situation, one must consider transitions between four different
statuses(open, non-sliding, and two different sliding directions).
One would like to be sure that such transitions are always uniquely
determined.  Unfortunately, the methods developed in this paper are not
sufficient to show this, so that this possible source of indeterminacy must
be investigated in the future.

This paper is organized as follows.  Sec.~\ref{synopsis} presents
an overview of the paper, detailing the questions posed in the
introduction, and sketching the results of the rest of the paper.  
Readers not wishing to savor the details
may read this section, and then skip directly to
Sec.~\ref{conclusion} for a discussion of the results.
The stiffness matrix for frictional disks is derived in
Sec.~\ref{QQD}, and includes a discussion of mechanism indeterminacy
in Sec.~\ref{mechindet}.  Sec.~\ref{statusindet} deals with
contact status indeterminacy by showing that there is a unique way
to choose contact statuses.

\section{Synopsis}
\label{synopsis}

\subsection{The Stiffness matrix}
\label{StiffnessIntro}

\begin{figure}
\centering
\includegraphics[width=0.5\columnwidth]{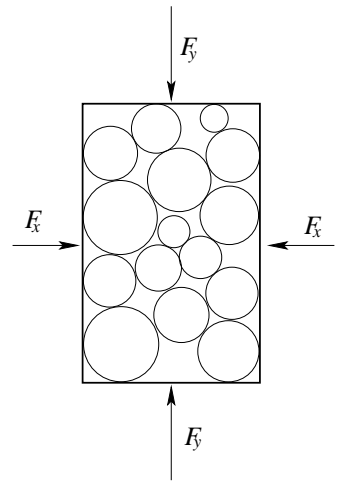}
\caption{Sketch of a biaxial test.  An assembly of two-dimensional
grains is confined by four walls.  Forces $F_x$ and $F_y$ are applied
to the vertical and horizontal walls.}
\label{biaxial}
\end{figure}

In this paper, we will deal with an assembly of disks,
interacting via Coulomb friction and subjected to a slowly
changing force.  As a concrete example,
consider a biaxial box,
where a granular sample composed of disks is enclosed in
a rectangular box of dimensions $L_x \times L_y$, with forces 
$F_x$ and $F_y$ exerted on the walls.  These forces vary
slowly with time, and one measures the resulting movement of the
walls.  A sketch of the biaxial box is shown in Fig.~\ref{biaxial}.

In Sec.~\ref{QQD}, we define the quasi-rigid limit precisely, and show that 
it leads to a piece-wise linear behavior of the packing.  Thus
time can be divided into intervals $[t_i,t_{i+1}]$ during which
the velocities of the particles are linearly related to the
change in forces:
\begin{equation}
\frac{d\mathbf{f}_\mathrm{ext}}{dt} = \mathbf{kv},
\label{PreviewStiffness}
\end{equation}
where $\mathbf{f}_\mathrm{ext}$ represents the external forces
($F_x$ and $F_y$ for the biaxial box), $\mathbf{v}$ contains
the velocities of the particles, and $\mathbf{k}$ is called
the \textsl{stiffness matrix}.

The motion is only \textsl{piece-wise} linear because the stiffness
matrix $\mathbf{k}$ depends on the contact status.  Whenever a contact
status changes, therefore, $\mathbf{k}$ must be modified.  Therefore,
the times $\{t_i\}$ which define the intervals of linearity
are the times when one or more contacts change status.

\subsection{Indeterminacy of mechanism}

What happens when the stiffness matrix has a zero eigenvalue?  In that
case, there exists $\mathbf{v}_*\ne0$ such that $\mathbf{kv}_*=0$.
Any multiple of $\mathbf{v}_*$ can be added to the solution of
Eq.~(\ref{PreviewStiffness}), and it would still be a solution.  Thus
it would seem that the theory is incapable of determining the amplitude
of $\mathbf{v}_*$.  

But it important to realize that one obtains Eq.~(\ref{PreviewStiffness})
under the assumption that the external
forces change on a time scale that is very long compared to the vibrations
in the granular packing.  The appearance of a zero eigenvalue corresponds
to a diverging time scale of vibration, and thus the assumptions leading
to Eq.~(\ref{PreviewStiffness}) are not met.  One must use instead
Newton's second law, and in this case, the amplitude of $\mathbf{v}_*$
can be determined.  If there is no interaction between the mechanism and
the external force ($\mathbf{f}_\mathrm{ext}\cdot\mathbf{v}_*=0$), then
the mechanism is decoupled from the other degrees of freedom, and
Eq.~(\ref{PreviewStiffness}) can still be used.

\subsection{Contact Status Indeterminacy}
\label{CoulombIntro}

Another source of indeterminacy may arise from the dependence of $\mathbf{k}$
on the contact status.  Each contact in the packing may
be sliding or non-sliding,
and each choice leads to a different stiffness matrix.  If there are $M$
contacts, there are $2^M$ ways to assign contact status, and thus
$2^M$ possible stiffness matrices, and $2^M$ different solutions to
Eq.~(\ref{PreviewStiffness}).  Which one is correct?  As stated in
the introduction, there is a unique solution if the contacts have
Coulomb friction, and slide only when necessary.  Coulomb
friction means that the condition
\begin{equation}
\tilde F \equiv \mu F_n - |F_t| \ge 0.
\label{Coulombcondition}
\end{equation}
must be obeyed at each contact.
Here, $F_n$ and $F_t$ are the normal and tangential components
of the contact force.  The constant $\mu$ is the Coulomb friction coefficient.

When we say that contacts should slide only when necessary, we mean that
they should slide only when they would violate Eq.~(\ref{Coulombcondition})
if they did not slide.  Since this rule places an important role in this
paper, we give it a name:
\begin{quote}
\textsl{Principle of minimum sliding}: A contact slides if, and only if,
remaining non-sliding would violate Eq.~(\ref{Coulombcondition}).
\end{quote}

\begin{figure}
\centering
\includegraphics[width=0.5\columnwidth]{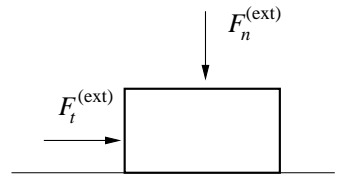}
\caption{A block pushed against a plane.}
\label{block}
\end{figure}

To illustrate this principle, let us consider
a block pushed against a plane with normal force $F_n^\mathrm{(ext)}$,
as shown in Fig.~\ref{block}.
A force tangential to the plane $F_t^\mathrm{(ext)}$ is also applied.
The contact between the block and the plane exerts a normal force
$F_n^\mathrm{(C)}$ and a tangential force $F_t^\mathrm{(C)}$.
Let these forces be directed opposite to the external ones, so
$F_{n,t}^\mathrm{(ext)}=F_{n,t}^\mathrm{(C)}$ indicates force
equilibrium.
Let us suppose that $F_n^\mathrm{(ext)}$ is fixed, while $F_t^\mathrm{(ext)}$
is slowly increased from zero.  As long as $F_t^\mathrm{(ext)} \le \mu
F_n^\mathrm{(ext)}$, we have $F_n^\mathrm{(C)}=F_n^\mathrm{(ext)}$
and $F_t^\mathrm{(C)}=F_n^\mathrm{(ext)}$, so
the block remains in place, and the contact
is non-sliding.  When $F_t^\mathrm{(ext)} > \mu
F_n^\mathrm{(ext)}$, the block must begin to slide, since the contact
cannot cancel the imposed tangential force without violating
Eq.~(\ref{Coulombcondition}).  The contact is now sliding and 
$F_t^\mathrm{(C)} = \mu F_n^{(ext)}$.  Now suppose
$F_t^\mathrm{(ext)}$ is decreased again, so that  $F_t^\mathrm{(ext)} \le \mu
F_n^\mathrm{(ext)}$.  The block will de-accelerate, and finally stop.
When the block stops, the principle of minimum sliding says that
the contact must become non-sliding.
Then the block will remain in place as long as 
$F_t^\mathrm{(ext)} \le \mu F_n^\mathrm{(ext)}$.
Note that if we did not apply the principle of minimum sliding, we would
obtain a nonsensical result: the tangential contact force would remain
constant at $\mu F_n^\mathrm{(ext)}$ and start to accelerate
the block.  The frictional forces would thus be doing work on the block,
which is a violation of the second law of thermodynamics.
Thus, there is nothing artificial about the principle of minimum sliding.
It is simply makes explicit what is needed to obtain
sensible results.  It also describes what is done in numerical simulations.

\begin{figure}
\centering
\includegraphics[width=0.5\columnwidth]{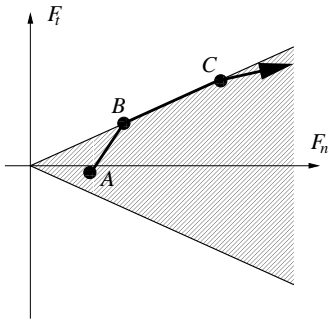}
\caption{The possible motion of a contact in the $(F_n,F_t)$ plane.
The Coulomb cone is shaded.  The contact begins at point $A$ within
the cone.  At $B$, it reaches the surface of the cone, becomes
sliding, and moves along the cone boundary.  At $C$ it becomes non-sliding
and moves into the cone interior.}
\label{cone}
\end{figure}

Now let us return to the problem of choosing the contact statuses
in a granular packing.  It is helpful to consider the contact
forces as points in the ($F_n,F_t)$ plane.
The set of forces $(F_n,F_t)$ that obey Eq.~(\ref{Coulombcondition}) is
shown in Fig.~\ref{cone} as a shaded region.  This set is
called the \textsl{Coulomb cone} because it forms a cone when plotted this way.

As the particles move, the contact forces change, and thus trace out continuous
trajectories in the $(F_n,F_t)$ plane.  These trajectories must of course
remain within the Coulomb cone.  For contacts in the interior of the Coulomb cone,
any motion is allowed (for short enough times), since there is no danger
they will leave the cone.  Therefore, all such contacts will be non-sliding
by the principle of minimum sliding.

Contacts whose forces lie on the boundary of the Coulomb cone
$|F_t|=\mu F_n$ are called \textsl{critical} contacts,
and must be handled carefully, since they may leave the Coulomb cone.
If they are sliding, they will stay on the boundary.  But as we saw above,
we must also allow them to leave this surface and enter back into the
interior of the Coulomb cone.  Therefore, each critical contact
could be sliding or non-sliding.

It seems that one could determine the status of the critical contacts 
simply by inspecting the particle velocities $\mathbf{v}$.  
These velocities determine the change in contact
forces, and one can easily determine if these changes would cause
a critical contact to return to the interior of the Coulomb cone or not.
However, changing a contact status also changes the matrix $\mathbf{k}$
and thus through Eq.~(\ref{PreviewStiffness}) the velocities $\mathbf{v}$.
These new velocities may require changing the status of other contacts,
provoking another re-calculation of $\mathbf{v}$, etc.  Thus it seems
one must use an iterative procedure.  One assigns the contact status in a
certain way, uses Eq.~(\ref{PreviewStiffness}) to calculate the particle
velocities $\mathbf{v}$, and then begins to check if these velocities
are consistent with the chosen statuses.
If an inconsistency is found, the status must be changed,
and the procedure begins again.  This procedure must be continued until
a solution is found.

In this paper, we are not concerned with this algorithm, but rather about
the uniqueness and existence of a solution.  Is it always possible to find
a solution?  Or are there many solutions?   Note that it is difficult
to investigate these questions numerically.  Even though we must only deal
with the critical contacts, we are often faced with situations where all
possibilities cannot be investigated.  For example, it is common to have
hundreds of critical contacts in numerical simulations involving thousands
of particles.  This means that the possible ways to choose the status cannot
even be numbered with 64-bit integers.   

Either the non-existence or non-uniqueness would bring up hard questions
about the quasi-rigid approach.  If there were sometimes 
no solutions, the theory could not be applied to those situations.
On the other hand, if the solution were not unique, the stiffness matrix,
combined with the principle of minimum sliding, would not be a
complete description of the system.  Some physical process must
decide between the different possibilities.  This unknown
process would have been left
out of the model, leading to indeterminacy in the same way that 
neglecting particle deformations leads to force indeterminacy.
We would then have to ask what that physical process could be. 
One possibility is sound waves.  As we show below, the quasi-static
assumption amounts to removing `fast' processes like sound waves.
When a contact changes status, there is probably a ``negotiation''
between the critical contacts, mediated by sound waves,
that establishes their status.  In the quasi-static limit, this period of 
negotiation becomes a single point in time, and it is assumed
that the principle of minimum sliding suffices to determine the
new status.  Non-uniqueness of the choice of contact status means
that the details of this negotiation must be taken into account.

\subsection{Uniqueness of the solution}
\label{ProofPreview}

We now sketch the proof that
that there is always one, and only one choice of contact status that 
satisfies the principle of minimum sliding everywhere in the packing.
We begin by defining some terms.  Let the \textsl{state} of a packing
be a way of assigning the status to all the critical contacts.
To each state belongs
a corresponding set of velocities $\mathbf{v}$, which can be calculated from
Eq.~(\ref{PreviewStiffness}).  A state is \textsl{locally consistent at
contact $\alpha$} if the principle of minimum sliding is obeyed at that
contact, and \textsl{locally inconsistent} otherwise.  We also refer to
the \textsl{consistency} of a contact, 
which means whether or not the principle of minimum
sliding is obeyed there or not.  A state is \textsl{globally consistent} if
it is locally consistent at all contacts.  We are thus concerned with
the existence and uniqueness of the globally consistent state.

The proof has two premises.  First,
we assume that all possible states lead to
a stable packing.  The packing is stable if
\begin{equation}
\mathbf{v}^T\mathbf{kv} > 0,
\label{PreviewStability}
\end{equation}
for a certain (large) class of relevant vectors $\mathbf{v}$.
The second premise is the observation 
that the left hand side of Eq.~(\ref{PreviewStiffness})
is not modified by the status of the contacts.  Thus if we consider
two different states $X$ and $Y$ for the global contact status,
one has
\begin{equation}
\frac{d\mathbf{f}_\mathrm{ext}}{dt} = \mathbf{k}^X \mathbf{v}^X =
  \mathbf{k}^Y \mathbf{v}^Y,
\label{PreviewSameSol}
\end{equation}
where $\mathbf{k}^X$ is the stiffness matrix obtained if one chooses
$X$, and $\mathbf{k}^Y$ is obtained by choosing $Y$.  The
corresponding velocities are $\mathbf{v}^X$ and $\mathbf{v}^Y$.

From Eqs.~(\ref{PreviewStability}) and (\ref{PreviewSameSol}) it is
possible to derive a series of inequalities, from which one may deduce
the following theorem:
\begin{quote}
\textsl{Status change theorem:}
If the status of any set of critical contacts changes,
the consistency of at least one of those contacts must also change.
\end{quote}
This statement is sufficient to prove both existence and uniqueness.

To show this, let us cast the theorem into
a different form.  A particular state corresponds
to a $M_c$-bit binary number, $S\in\{0,1\}^{M_c}$, 
with each bit corresponding to the status
of a single critical contact, and $M_c$ is the number of critical contacts.  
For concreteness, let us say $S_\alpha=1$
if contact $\alpha$ is sliding, and $S_\alpha=0$ if it
is non-sliding.   To check the consistency of a given state $S$
of contact status, we would construct the corresponding stiffness
matrix $\mathbf{k}^S$, solve Eq.~(\ref{PreviewStiffness}) for
$\mathbf{v}^S$, and then check for consistency at each critical contact.
The result of this procedure can be represented by a second
$M_c$-bit binary number $C=\mathcal{C}(S)$, where each bit gives the
consistency of a critical contact, i.e. $C_\alpha=1$
if contact $\alpha$ is consistent, and $C_\alpha=0$ otherwise.
Now, the above theorem can be stated as follows: Changing any number
of bits of $S$ causes at least one of the corresponding bits in
$C$ to change.  This means that no two different values of $S$ can
lead to the same value of $C$.  There are $2^{M_c}$ possible choices
for $S$, and $2^{M_c}$ possible values for $C$, so each possible
value of $C$ must be associated with a unique value of $S$.  This
applies also to $C=111\ldots1$, corresponding to global consistency.

This result can be elegantly stated using a more mathematical language.
The process of determining the consistency of a state
defines a mapping $\mathcal{C}$ of one
$M_c$-bit binary number to another:
\begin{eqnarray}
\mathcal{C} : \{0,1\}^{M_c} &\to& \{0,1\}^{M_c}\cr
S & \mapsto & \mathcal{C}(S).
\end{eqnarray}
Now the status change theorem becomes simply: \textsl{the mapping
$\mathcal{C}$ is bijective}.  This proves the existence and uniqueness 
of the consistent choice since $\mathcal{C}$ is a mapping from
$\{0,1\}^{M_c}$ onto itself.

\section{The stiffness matrix}
\label{QQD}

In this section, we present a derivation of Eq.~(\ref{PreviewStiffness}),
where the motion of the particles is related to the change in applied
force by the stiffness matrix.  Several related derivations have already
been published \cite{Roux,Kuhn,Baginew,Miehe}.  The formulation presented
here is distinguished from these other works in several ways.  First, it
incorporates sliding contacts, which
is necessary to consider the uniqueness of the
globally consistent contact status.  The second difference is that
Eq.~(\ref{PreviewStiffness}) is shown to be a certain limit of
Newton's second law.  This is essential to resolving the question of
mechanism indeterminacy.  On the other hand, the simplest possible form
of the grains -- disks in two dimensions -- is assumed.

We first describe how interactions between the grains are modeled.
We then assemble the quantities introduced here into vectors and
matrices that describe all particles in the assembly.  We then
insert these results into Newton's second law to obtain equations
for the motion of the grains.  Then a limit of these equations is
taken, leading to Eq.~(\ref{PreviewStiffness}).  A discussion
of mechanism indeterminacy completes this section.

\subsection{Particle interaction model}
\label{ParticleModel}

We suppose that the grains interact through cohesionless repulsion
and Coulomb friction.  We adopt the convention that a positive 
normal contact force $F_n$ corresponds to repulsion.  Thus the absence
of cohesion requires
\begin{equation}
F_n \ge 0.
\label{NoCohesion}
\end{equation}
Recall that Coulomb friction means that Eq.~(\ref{Coulombcondition}) is
obeyed.

When two
grains first touch, two springs are created, one in the tangential and
the other in the normal direction.  The springs obey Hooke's law 
so that the normal and tangential contact
forces $F_n$, $F_t$ are proportional to the spring elongations $D_n$, $D_t$.
To this restoring force, we add a linear damping to model the dissipation
of energy:
\begin{equation}
F_n = -K_n D_n - \Gamma_n V_n, \quad F_t = -K_t D_t - \Gamma_t V_t,
\label{Feqn}
\end{equation}
where $K_n$ and $K_t$ are the spring constants, $\Gamma_n$ and $\Gamma_t$
are viscous damping coefficients, and $V_n$ and $V_t$ are the
the normal and tangential relative velocities.  
Here, $D_n<0$ is interpreted as an
overlap. 

The springs are stretched by the relative motion of the particles,
as long this does not violate Eq.~(\ref{Coulombcondition})
or Eq.~(\ref{NoCohesion}).  When the contact is in the interior
of the Coulomb cone, any motion is possible, so one has
\begin{equation}
\frac{dD_n}{dt} = V_n, \quad
\frac{dD_t}{dt} = V_t,
\label{DeqnNS}
\end{equation}
where $V_n$ and $V_t$ are just the relative velocities at the point
of contact:
\begin{eqnarray}
V_n &=& (\vec v_i - \vec v_j)\cdot \hat n \hat n,\cr
V_t &=& (\vec v_i - \vec v_j)\cdot \hat t \hat t 
  + r_i \omega_i + r_j \omega_j,
\label{Veqn}
\end{eqnarray}
where $\vec v_i$, $\omega_i$, and $r_i$ are the velocity, angular
velocity and radius of particle $i$, and $i$ and $j$ label the
touching particles.  The vectors $\hat n$ and $\hat t$ are
unit vectors pointing in the normal and tangential directions,
respectively.
Throughout this paper, capital letters
indicate quantities concerning contacts, and small letters
quantities concerning particles.

Now let us consider how to handle sliding contacts.  It is helpful
to define
\begin{equation}
\tilde V = \mu \frac{K_n}{K_t} V_n + V_t \sgn D_t.
\label{defVs}
\end{equation}
Note that if the contact is non-sliding,
\begin{equation}
\frac{d\tilde F}{dt} = -(K_t+\Gamma_t) \tilde V.
\label{FsVs}
\end{equation}
where $\tilde F$ is defined in Eq.~(\ref{Coulombcondition}).  
For contacts on the boundary of the Coulomb cone, we have $\tilde F=0$.
The sign of $\tilde V$ determines whether such contacts leave or remain
within the Coulomb cone when made non-sliding.  If $\tilde V<0$, the
contact will move into the interior of the Coulomb cone ($\tilde F>0$).
If $\tilde V>0$, the point would leave the Coulomb cone.
The principle of minimum sliding can thus be reformulated:
\begin{quote}
\textsl{Principle of minimum sliding:}
The status `sliding' is consistent if, and only if, $\tilde V > 0$,
where $\tilde V$ is defined in Eq.~(\ref{defVs}).
\end{quote}

When a contact slides,
Eq.~(\ref{Feqn}) is still valid, but we set $\Gamma_t=0$
and constrain the spring elongations to change so that $\tilde F=0$.
This can be accomplished
if we use the first equation in Eq.~(\ref{DeqnNS}) but replace the second
with
\begin{equation}
\frac{dD_t}{dt} = -\left[\mu\frac{K_n}{K_t} \sgn D_t\right] V_n.
\label{DeqnS}
\end{equation}

Once the contact forces are known, the net force $\vec f$ and torque
$\tau$ on each particle can be computed:
\begin{eqnarray}
\vec f &=& \sum_\alpha F_{\alpha,n} \hat n_\alpha 
   + F_{\alpha,t} \hat t_\alpha,\cr
\tau &=& r\sum_\alpha F_{\alpha,t},
\label{feqn}
\end{eqnarray}
where the sums are taken over all the contacts that the concerned particle
makes with its neighbors, and $r$ is its radius.

\subsection{Matrix formulation}

It is useful to consider the proceeding equations in matrix
form.  To do so, we must gather the various quantities into vectors.
To begin with, we can group the force and torque exerted on particle
$i$ into a vector $\underline{f}_i$, and the contact forces exerted by 
a contact $\alpha$ into a vector $\underline{F}_\alpha$:
\begin{equation}
\underline{f}_i = \left( \begin{array}c
f_{i,x} \\ f_{i,y} \\ \tau_i/r_i
\end{array} \right), \quad
\underline{F}_\alpha = \left( \begin{array}c
F_{\alpha,n} \\ F_{\alpha,t}
\end{array} \right).
\end{equation}
It is often convenient to group these vectors together into high-dimensional
quantities concerning all the particles or contacts in the packing:
\begin{equation}
\mathbf{f} = \left( \begin{array}c
	\underline{f}_1 \\ \underline{f}_2 \\ \vdots \\ \underline{f}_N
\end{array} \right)
 = \left( \begin{array}c
   f_{1,x} \\ f_{1,y} \\ \tau_1/r_1 \\ \vdots \\ 
   f_{N,x} \\ f_{N,y} \\ \tau_N/r_N
\end{array} \right),
\label{matfdef}
\end{equation}
and
\begin{equation}
\mathbf{F} = \left( \begin{array}{c}
	\underline{F}_1 \\ \underline{F}_2 \\ \vdots \\ \underline{F}_N
\end{array} \right) 
 = \left( \begin{array}{c}
  F_{1,n} \\ F_{1,t} \\ \vdots \\ F_{M,n} \\ F_{M,t} 
\end{array} \right).
\label{matFdef}
\end{equation}
Here $N$ is the number of bodies 
whose motion must be considered, and $M$ is the number of contacts
between these bodies.  In these equations, and throughout this paper,
boldface vectors will denote
quantities concerning all contacts or particles (i.e. vectors
in contact or particle space), whereas underscores
indicate quantities associated with a single particle or contact.

Eq.~(\ref{feqn}) can now be written
\begin{equation}
\underline{f}_i 
  = \sum_{\alpha=1}^M \underline{c}_{i\alpha} \underline{F}_\alpha,
\label{fvec}
\end{equation}
where $\underline{c}_{i\alpha}$ is a $3 \times 2$ matrix
\begin{equation}
\underline{c}_{i\alpha} = \left( \begin{array}{cc}
\chi_{i\alpha} \hat n_{\alpha x} & \chi_{i\alpha} \hat t_{\alpha x} \\
\chi_{i\alpha} \hat n_{\alpha y} & \chi_{i\alpha} \hat t_{\alpha y} \\
0 & |\chi_{i\alpha}| \end{array} \right).
\label{cmatrix}
\end{equation}
This gives the contribution of
contact $\alpha$ to the force exerted on particle $i$.  The symbol
$\chi_{i\alpha}$ is defined as
\begin{equation}
\chi_{i\alpha} = \left\{
   \begin{array}{cl}
	1 & \mbox{if particle }i\mbox{ is first in contact }\alpha,\\
	-1 & \mbox{if particle }i\mbox{ is second in contact }\alpha,\\
	0 & \mbox{if particle }i\mbox{ does not participate in contact }\alpha.
   \end{array} \right.
\end{equation}
If a particle is ``first'' in contact $\alpha$, that means that the
contact exerts a normal force $F_{n,\alpha}\hat n_\alpha$ on it.
If it is ``second'' in contact $\alpha$, a normal force
$-F_{n,\alpha}\hat n_\alpha$ is exerted on it.
For each contact between two grains, one element of $\chi$ is $1$, 
and another is $-1$.  $\chi$ is also called the incidence matrix.

Eq.~(\ref{fvec}) holds for each particle ($i=1\dots N$).  All of these
equations can be written compactly using the definitions in
Eqs.~(\ref{matfdef}) and (\ref{matFdef}):
\begin{equation}
\mathbf{f} = \mathbf{cF}.
\label{fmat}
\end{equation}
The $3N\times2M$ matrix $\mathbf{c}$ can be constructed assembling
an $N\times M$ array of the $\underline{c}_{i\alpha}$.

One can consider Eq.~(\ref{fmat}) as an equation for the unknown
contact forces $\mathbf{F}$.  However, one
almost always has $2M>3N$, meaning that $\mathbf{c}$ has at least
$2M-3N$ linearly independent
null eigenvectors.  Therefore, Eq.~(\ref{fmat}) does not
have a unique solution.  This is the force indeterminacy problem
discussed elsewhere in the literature 
\cite{indet,FNensemble,one,Radjai96,Wolf}.

We now continue by gathering the other quantities introduced in
Sec.~\ref{ParticleModel} into vectors.
Eq.~(\ref{Feqn}) can be written
\begin{equation}
\underline{F}_\alpha = -\underline{K}_\alpha\underline{D}_\alpha
 -\underline{\Gamma}_\alpha\underline{V}_\alpha,
\mbox{ or }
\mathbf{F} = -\mathbf{KD} - \mathbf{\Gamma V},
\label{Fmat}
\end{equation}
where
\begin{equation}
\underline{K}_\alpha = \left( \begin{array}{cc}
K_n & 0 \\ 0 & K_t 
\end{array} \right),
\quad
\underline{\Gamma}_\alpha = \left( \begin{array}{cc}
\Gamma_n & 0 \\ 0 & \Gamma_t 
\end{array} \right),
\label{Kdef}
\end{equation}
and $\mathbf{K}$ and $\mathbf{\Gamma}$ are $2M\times 2M$ diagonal matrices
containing the $\underline{K}_\alpha$ or the $\underline{\Gamma}_\alpha$
on the diagonal.

Eq.~(\ref{Veqn}) can be written as
\begin{equation}
\underline{V}_\alpha = \sum_{i=1}^N(\underline{c}_{i\alpha})^T \underline{v}_i,
\mbox{ or }
\mathbf{V} = \mathbf{c}^T \mathbf{v},
\label{Vmat}
\end{equation}
where $\mathbf{c}^T$ is the transpose of $\mathbf{c}$ \cite{Roux}.  
Since the dimension of $\mathbf{V}$ is larger than
that of $\mathbf{v}$,
Eq.~(\ref{Vmat}) places restrictions on $\mathbf{V}$.  Not
every vector $\mathbf{V}\in\mathbb{R}^{2M}$ is allowed, but only those
vectors in the range of $\mathbf{c}^T$.  Physically, this means that
not every relative motion is possible, but only those that
can be generated by moving and rotating the particles.  The dimension
of the range of $\mathbf{c}^T$ is at most $3N$.  There are thus $2M-3N$ 
dimensions in $\mathbb{R}^{2M}$ that are inaccessible.  These
$2M-3N$ dimensions are precisely the null space of $\mathbf{c}$ \cite{Roux}.

Finally, the relation between $\mathbf{D}$ and $\mathbf{V}$ in
Eqs.~(\ref{DeqnNS}) and (\ref{DeqnS}) requires careful treatment,
due to the different possible contact statuses.  Let $\mathbb{S}$
be the set of sliding contacts.
We define a $2\times2$ matrix $\underline{S}_\alpha$ 
that depends on the status of contact $\alpha$.
If $\alpha\not\in\mathbb{S}$, $\underline{S}_\alpha=\underline{1}$,
if $\alpha\in\mathbb{S}$:
\begin{equation}
\underline{S}_\alpha = \left(\begin{array}{cc} 1 & 0 \\
- \mu \frac{K_n}{K_t}\,\sgn D_t & 0 \end{array} \right).
\label{Ssliding}
\end{equation}
Now the relation between $\mathbf{v}$ and $\mathbf{D}$ can be written:
\begin{equation}
\frac{d\underline{D}_\alpha}{dt} 
	= \underline{S}_\alpha\underline{V}_\alpha,
\mbox{ or }
\frac{d\mathbf{D}}{dt} = \mathbf{S}(\mathbb{S})\mathbf{V}.
\label{Dmat}
\end{equation}
$\mathbf{S}(\mathbb{S})$ is a block diagonal matrix, with the 
$\underline{S}_\alpha$ on the diagonal.  It is a function of
$\mathbb{S}$, as indicated.

Note that $\sgn D_t$ in Eq.~(\ref{Ssliding}) is a constant.
In order for $\sgn D_t$ to change, the contact must cross
the $F_n$ axis in Fig.~\ref{cone}.  This can only happen
if the contact passes through the interior of the Coulomb cone.
In that case, the contact would be non-sliding, and Eq.~(\ref{Ssliding})
would not be applied.  The exception to this occurs when a contact
approaches the origin.  This brings up the question of what happens
when a contact opens or closes.  We are not dealing with that problem
in this paper.

\begin{table}
\begin{tabular}{rclcc}
\hline
\multicolumn{3}{c}{Linear Relation} &\multicolumn{2}{c}{Equation}\\
\hline
force on particles&$\propto$&contact forces & $\mathbf{f}=\mathbf{cF}$&
(\ref{fmat})\\
contact forces&$\propto$&spring lengths& $\mathbf{F}=-\mathbf{KD}$&
(\ref{Fmat})\\
change in spring length&$\propto$&relative motion&
$\mathbf{\dot D}=\mathbf{SV}$&(\ref{Dmat})\\
relative motion&$\propto$&particle motion& 
$\mathbf{V} = \mathbf{c}^T \mathbf{v}$&(\ref{Vmat})\\
\hline
\end{tabular}
\caption{Summary of stiffness matrix derivation as a chain of
linear relations.  The symbol `$\propto$'
is used to mean ``is linearly related to''.}
\label{summary}
\end{table}

Our derivation of the global stiffness matrix is summarized in
Table.~\ref{summary}.  It rests on a chain of linear relations
that can be established (or assumed) between the various quantities.

\subsection{Equations of motion}

At this point, most derivations of the stiffness matrix proceed directly
to force equilibrium, and assume that the net forces $\mathbf{f}$
exerted on each particle are balanced by some externally imposed load
$\mathbf{f}_\mathrm{ext}$.  We take a longer route that gives more
insight into the situations where force equilibrium does not hold.
We begin with Newton's second law, which relates the accelerations
of the particles to the forces exerted on them:
\begin{equation}
\mathbf{m} \frac{d\mathbf{v}}{dt} = \mathbf{f}+\mathbf{f}_\mathrm{ext}.
\label{Newton}
\end{equation}
Here, $\mathbf{m}$ is a diagonal matrix containing the masses and momenta
of inertia of all the grains.  We could also write
\begin{equation}
\underline{m}_{i} \frac{d\underline{v}_i}{dt} = \underline{f}_i +
\underline{f}_{\mathrm{ext},i}.
\end{equation}
with
\begin{equation}
\underline{m}_i = \left( \begin{array}{ccc}
m_i & 0 & 0 \\ 0 & m_i & 0 \\ 0 & 0 & I_i/r_i^2
\end{array} \right),
\end{equation}
where $m_i$ is the mass of particle $i$ and $I_i$ is its moment of inertia.

Combining Eq.~(\ref{fmat}), (\ref{Fmat}) and (\ref{Newton}) gives
\begin{equation}
\mathbf{m} \frac{d\mathbf{v}}{dt} = -\mathbf{cKD}+\mathbf{f}_\mathrm{ext}.
\label{Newton2}
\end{equation}
This equation can be differentiated once with respect to time, and
Eqs.~(\ref{Vmat}) and (\ref{Dmat}) can be used to obtain
\begin{equation}
\mathbf{m} \frac{d^2\mathbf{v}}{dt^2} = -\mathbf{cKSc}^T\mathbf{v} -
\frac{d\mathbf{c}}{dt}\mathbf{KD}
- \mathbf{c}^T\mathbf{\Gamma c} \frac{d\mathbf{v}}{dt}
+\frac{d\mathbf{f}_\mathrm{ext}}{dt}.
\end{equation}
The  combination $\mathbf{cKSc}^T$ appears often, so we define
$\mathbf{k}=\mathbf{cKSc}^T$ and write
\begin{equation}
\mathbf{m} \frac{d^2\mathbf{v}}{dt^2} = -\mathbf{kv} -
\frac{d\mathbf{c}}{dt}\mathbf{KD}
- \mathbf{c}^T\mathbf{\Gamma c} \frac{d\mathbf{v}}{dt}
+\frac{d\mathbf{f}_\mathrm{ext}}{dt}.
\label{fulleq}
\end{equation}
This equation gives the full motion, without approximation, of the
disks.  Such an equation is solved numerically in the ``molecular 
dynamics'' simulation method.   On the left hand side is the mass
times the acceleration (differentiated by time), and on the
right hand side are the forces exerted on the particles (also
differentiated by time).

\subsection{Quasi-static balance}
\label{Qbalance}

If one makes the quasi-rigid and quasi-static assumptions, then
two terms dominate in Eq.~(\ref{fulleq}).  The quasi-rigid assumption
means that the hardness of the particles is assumed to be much
greater than the confining pressure, and  
the quasi-static assumption is that the external forces 
$\mathbf{f}_\mathrm{ext}$ change much more slowly than
any timescale associated with the contact forces.  We describe
the consequences of each of these assumptions below.

Let us begin with the quasi-rigid assumption.  We will
compare the first two terms on the right hand side of Eq.~(\ref{fulleq}).
From Eq.~(\ref{cmatrix}), we see that $\frac{d\mathbf{c}}{dt}$ will
be proportional to $\frac{d\hat n}{dt}$.  Carrying out this differentiation:
\begin{equation}
\frac{d\hat n}{dt} = \frac{d}{dt}
   \left( \frac{\vec x_i - \vec x_j}{\left|\vec x_i - \vec x_j \right|} \right)
   = \frac{\vec v_i-\vec v_j}{\left|\vec x_i - \vec x_j \right|}
      \cdot \hat t \hat t
 \sim O\left(\frac{V}{R}\right).
\end{equation}
Here, $R$ is a typical particle radius, and $V$ a typical relative
velocity.  Now we can estimate the sizes of the first two terms
on the right hand side of Eq.~(\ref{fulleq}):
\begin{equation}
\mathbf{kv} = \mathbf{cKSV} \sim O\left(KV\right), \quad
\frac{d\mathbf{c}}{dt}\mathbf{KD} \sim O\left(KV\frac{D}{R}\right),
\end{equation}
where we have used $\mathbf{c}\sim \mathbf{S} \sim O(1)$.
The quasi-rigid assumption implies
that the deformations $D$ are much smaller than
a typical particle radius $R$, or $D/R \ll 1$.  Thus the second
term on the right hand side is much smaller than the first,
and Eq.~(\ref{fulleq}) becomes
\begin{equation}
\mathbf{m} \frac{d^2\mathbf{v}}{dt^2} = -\mathbf{kv} 
- \mathbf{c}^T\mathbf{\Gamma c} \frac{d\mathbf{v}}{dt}
+\frac{d\mathbf{f}_\mathrm{ext}}{dt}.
\label{quasirigid}
\end{equation}
The term that we have just neglected is called the ``geometric stiffness''
\cite{Kuhn}.  It can again become important in some situations.  One
example will be given in Sec.~\ref{mechindet}.

Now let us proceed to the quasi-static assumption. 
As stated above, this means that the external force $\mathbf{f}_\mathrm{ext}$
is assumed to change much more slowly than any time scale associated
with the vibrations in the granular assembly.
To express the separation of time scales, we write
\begin{equation}
\frac{d}{dt} = \frac{d}{dt_0} + \epsilon \frac{d}{dt_1},
\end{equation}
where the variable $t_0$ measures a fast time scale, and $t_1$ a
long time scale.  In the problem at hand, the fast time scale describes
vibrations of the packing, and the slow time scale is given by the
change in the external load $\mathbf{f}_\mathrm{ext}$.
The presence of $\epsilon\ll1$ before the derivative
with respect to $t_1$ shows that these derivatives are small,
as $t_1$ measures slow changes.

The assumption of quasi-staticity can be expressed by saying that the
particle positions $\mathbf{x}$ depend on both $t_0$ and $t_1$, but
the external force $\mathbf{f}_\mathrm{ext}$ depends only on the
slow time $t_1$:
\begin{equation}
\mathbf{x} = \mathbf{x}_0(t_0) + \mathbf{x}_1(t_1), \quad
\mathbf{f}_\mathrm{ext} = \mathbf{f}_\mathrm{ext}(t_1).
\end{equation}
After differentiation by time, we have
\begin{equation}
\mathbf{v} = \mathbf{v}_0 + \epsilon \mathbf{v}_1.
\end{equation}
and the $O(1)$ terms of Eq.~(\ref{fulleq}) are
\begin{equation}
\mathbf{m} \frac{d^2\mathbf{v}_0}{dt_0^2} = -\mathbf{kv}_0 -
  \mathbf{c}\mathbf{\Gamma c}^T \frac{d\mathbf{v}_0}{dt_0}.
\label{oscillate}
\end{equation}
This equation resembles that of a damped, harmonic oscillator, with three
differences.  First, it is an equation for the velocity, not the position.
Second, it is a vector equation, not a scalar one.  Finally,
$\mathbf{k}$ and $\mathbf{\Gamma}$ depend on contact status.  Nevertheless,
it has the same properties as a damped, harmonic oscillator.  
The matrix $\mathbf{m}$ is positive definite and 
$\mathbf{c}^T\mathbf{\Gamma c}$ is non-negative definite,
as $\mathbf{m}$ and $\mathbf{\Gamma}$ are both diagonal
matrices with positive or non-negative entries.  
The stiffness matrix $\mathbf{k}$,
however is not necessarily positive definite.  If there are vectors
$\mathbf{v}\ne\mathbf{0}$ such that $\mathbf{v}^T\mathbf{kv}<0$, then
$\mathbf{k}$ acts like a negative number in Eq.~(\ref{oscillate}),
and $\mathbf{v}_0$ grows exponentially on a short time scale.
This is ``motion through an instability'' \cite{one}.  
Physically, the external forces push the particles \textsl{away}
from the positions they must occupy in order to be in equilibrium
\cite{Baginew}.  In this
case, one does not obtain the quasi-static balance; rather the packing
is unstable and is set in motion.

On the other hand, if $\mathbf{v}^T\mathbf{kv}>0$ 
then $\mathbf{k}$ acts like a positive number in Eq.~(\ref{oscillate}),
and $\mathbf{v}$ undergoes damped oscillations.  In this case
$\mathbf{v}_0\to0$ as $t_0\to\infty$.  One can then examine
the $O(\epsilon)$ equations assuming $\mathbf{v}_0=0$.  (The situation
when $\mathbf{v}^T\mathbf{kv}=0$ is more complicated, and will
be discussed in the next section.)  Keeping the $O(\epsilon)$
terms of Eq.~(\ref{Newton}) gives
\begin{equation}
\frac{d\mathbf{f}_\mathrm{ext}}{dt} = \mathbf{kv}_1.
\label{apap}
\end{equation}
This is the quasi-static balance, and the same as 
Eq.~(\ref{PreviewStiffness}).  

\subsection{Mechanism indeterminacy}
\label{mechindet}

Now let us consider mechanism indeterminacy.  This occurs
when $\mathbf{k}$ has null
eigenvalues, i.e., when $\mathbf{kv}_*=0$ for $\mathbf{v}_*\ne0$.   
The amplitude of these motions cannot be determined from Eq.~(\ref{apap}),
thus they appear to be a source of indeterminacy, just as null eigenvalues
of $\mathbf{c}$ [see Eq.~(\ref{fmat})] indicate force indeterminacy in 
rigid particle theories.  But there is one crucial difference between
Eq.~(\ref{fmat}) and Eq.~(\ref{apap}): the matrix $\mathbf{k}$ in
Eq.~(\ref{apap}) is a square matrix, so if it has null eigenvalue,
its range has a lower dimension than the left hand side of the
equation.  Thus there are external loads for which Eq.~(\ref{apap}) has
no solution.  On the other hand, the range of $\mathbf{c}$ can have
the same dimension as the right hand side of Eq.~(\ref{fmat}), even
when $\mathbf{c}$ has many null eigenvalues.

What happens when Eq.~(\ref{apap}) has no solution?  In that case,
the quasi-static assumption used to derive Eq.~(\ref{apap}) is invalid. 
This is because a null eigenvalue corresponds to a diverging period
of vibration in the packing.  Thus one cannot assume that the force
is changing on a time scale much greater than the period of
vibration.  Therefore, one
must consider Eq.~(\ref{quasirigid}) without assuming any separation
of time scales.  If one considers only the eigenvector $\mathbf{v}_*$,
Eq.~(\ref{quasirigid}) can be integrated once to give
\begin{equation}
\mathbf{m} \frac{d\mathbf{v}_*}{dt} = 
- \mathbf{c}^T\mathbf{\Gamma c} \mathbf{v}_*
+\mathbf{f}_\mathrm{ext}.
\end{equation}
If the damping can be neglected, then 
$|\mathbf{v}_*| \propto t |\mathbf{f}_\mathrm{ext}|/|\mathbf{m}|$.
This is ``motion through a mechanism'' \cite{one}.  The growth
of the velocity with time is much more gentle than for motion
through an instability that was discussed in the previous section.

However, Eq.~(\ref{apap}) can have a solution even when $\mathbf{k}$
has null eigenvalues.  In this case, the imposed force does not
excite the mechanisms.  The mechanisms are irrelevant to the evolution
of the system, and thus the derivation of Eq.~(\ref{apap}) is again
valid.  Such irrelevant mechanisms are quite common.  For example,
if one constructs the stiffness matrix of the biaxial box discussed
in Sec.~\ref{StiffnessIntro}, it is convenient 
to consider the walls as particles.  Then, the stiffness matrix 
automatically has three null eigenvalues: two associated with the
translation of the whole apparatus and one associated with its rotation.
As long as the forces on opposite walls are equal, these modes will not
be excited, and they are irrelevant.  
Another example is a ``rattler'' - a grain which
has no contacts.  Each one of its degrees of freedom yields a zero eigenvalue
of $\mathbf{k}$, but it can be removed from the system without changing
the quasi-static behavior.

Another thing that can happen when a mechanism is present is that
motion can be stabilized or destabilized by the geometric stiffness
$\frac{d\mathbf{c}}{dt}\mathbf{KD}$ in Eq.~(\ref{fulleq}), which
was neglected in the quasi-rigid limit.  When there is a mechanism,
the particle displacements are no longer required to be small,
so this term needs to be considered.  An example is when two elliptical
particles are pressed together \cite{Kuhn}.  If the particles are
circular, there is a mechanism: the two particles can roll like bearings
relative to one another.  If the particles are made elliptical,
the neglected term $\frac{d\mathbf{c}}{dt}\mathbf{KD}$ must be considered
to predict the behavior.

Thus, null eigenvalues of $\mathbf{k}$ can always be put into one of
two classes.  If Eq.~(\ref{apap}) has no solution, the null eigenvalue
signals the collapse of the packing, and the quasi-static assumption
fails, and Eq.~(\ref{quasirigid}) must be used to predict the motion
of the grains.  On the other hand, if Eq.~(\ref{apap}) still has a solution,
even though the range of $\mathbf{k}$ has been reduced in dimension,
the null eigenvalue corresponds to a degree of freedom that is irrelevant.
Null eigenvalues appear to cause indeterminacy only because Eq.~(\ref{apap})
is considered as the most fundamental equation.  However, Eq.~(\ref{apap}) is
in fact an approximation to Eq.~(\ref{quasirigid}). 

\section{Contact Status Indeterminacy}
\label{statusindet}

In this section, we consider contact status indeterminacy.  We will
show that only one state does not lead to a violation of the principle 
of minimum sliding at one or more contacts.  To do so, we must compare
the stiffness matrices of the different states.  We begin by presenting
the two hypotheses needed for the proof: first, that all possible states
are stable, and second that the applied load is independent of the 
state.  Then we consider an example where only two contacts are
critical, and show that the consistent state always exists and is unique.
Then we consider the general case with an arbitrary number of critical
contacts.

\subsection{Conditions needed to show uniqueness}
\label{premise}

\subsubsection{The stability condition}

A packing is stable if the
quadratic form $Q=\mathbf{v}^T\mathbf{kv}$ is positive 
\cite{one,Kuhn,Baginew}:
\begin{equation}
Q(\mathbf{v},\mathbb{S}) = \mathbf{v}^T\mathbf{k}(\mathbb{S})\mathbf{v}>0,
\label{stability}
\end{equation}
where $\mathbb{S}$ is the set of contacts that are sliding.
As we saw in Sec.~\ref{Qbalance}, the motion cannot be assumed to be
quasi-static when $Q\le0$. 

The quadratic form plays an important role in this paper, so we will
discuss how it can be calculated.  If we recall the definition
$\mathbf{k} =\mathbf{cKSc}^T$, and group factors in a suggestive way,
we have
\begin{equation}
Q(\mathbf{v},\mathbb{S})=
   [\mathbf{v}^T\mathbf{c}]^T\mathbf{KS}[\mathbf{c}^T\mathbf{v}].
= \mathbf{V}^T [\mathbf{KS}] \mathbf{V}
\end{equation}
The matrix $\mathbf{KS}$ is block diagonal, with each block corresponding
to a contact.  Thus the $Q(\mathbf{v},\mathbb{S})$ reveals itself
to be simply a sum over contacts:
\begin{eqnarray}
Q(\mathbf{v},\mathbb{S})&=& \sum_{\alpha=1}^M \underline{V}_\alpha^T
\underline{K}_\alpha \underline{S}_\alpha 
\underline{V}_\alpha,\cr
&=& \sum_{\alpha\not\in\mathbb{S}} Q_\alpha^\mathrm{(NS)} + 
 \sum_{\alpha\in\mathbb{S}} Q_\alpha^\mathrm{(S)}, 
\end{eqnarray}
where $Q_\alpha^\mathrm{(NS)}$ is the contribution of contact $\alpha$
if it is non-sliding, and $Q_\alpha^\mathrm{(S)}$ is its contribution
if it is sliding.  Using Eqs.~(\ref{Kdef}) and (\ref{Ssliding}), we have 
\begin{eqnarray}
Q^\mathrm{NS}_\alpha(\mathbf{v}) &=& K_n V_{n,\alpha}^2 + K_t V_{t,\alpha}^2,
\label{QNS}\\
Q_\alpha^\mathrm{(S)}(\mathbf{v}) &=& K_nV_{n,\alpha}^2 - \mu K_n V_{t,\alpha} 
  V_{n,\alpha} \sgn D_{t,\alpha}.
\label{QS1}
\end{eqnarray}
In the following, it is useful to use Eq.~(\ref{defVs}) and replace
$V_{n,\alpha}$ with $\tilde V_\alpha$.  Eq.~(\ref{QS1}) becomes
\begin{equation}
Q_\alpha^\mathrm{(S)}(\mathbf{v}) = Q_\alpha^\mathrm{(NS)}(\mathbf{v})
	- K_t V_{t,\alpha} \tilde V_\alpha \sgn D_{t,\alpha}.
\label{QS2}
\end{equation}
Now let us define
\begin{equation}
\hat F_\alpha = K_t V_{t,\alpha} \sgn D_{t,\alpha},
\end{equation}
so that Eq.~(\ref{QS2}) becomes
\begin{equation}
Q_\alpha^\mathrm{(S)}(\mathbf{v}) = Q_\alpha^\mathrm{(NS)}
	- \hat F_\alpha \tilde V_\alpha.
\label{QS}
\end{equation}
Therefore, the stability condition Eq.~(\ref{stability}) is
\begin{equation}
Q(\mathbf{v},\mathbb{S}) = Q(\mathbf{v},\emptyset) 
  - \sum_{\alpha\in\mathbb{S}} \hat F_\alpha \tilde V_\alpha > 0. 
\label{SlidingNegative}
\end{equation}
Note that $Q(\mathbf{v},\emptyset)>0$, because the contribution
of each contact must be positive.  This means the only way to
obtain an unstable packing is for the sliding contacts to make
large and negative contributions to $Q$.

In the following, it will be necessary to compare $Q$ for different
states.  If the sliding contacts present in a given state are
divided into two disjoint sets $\mathbb{S}_1$
and $\mathbb{S}_2$ ($\mathbb{S}_1\cap\mathbb{S}_2=\emptyset$), then
\begin{equation}
Q(\mathbf{v},\mathbb{S}_1\cup\mathbb{S}_2) = Q(\mathbf{v},\mathbb{S}_1) 
  - \sum_{\alpha\in\mathbb{S}_2} \hat F_\alpha \tilde V_\alpha, 
\label{stability2}
\end{equation}

\subsubsection{The independent load condition}

\begin{table}
\begin{tabular}{c|cccc}
&\multicolumn{4}{c}{-- Status --}\\
State&$\quad\mathbb{X}\quad$&$\quad\mathbb{Y}\quad$&
$\quad\mathbb{S}\quad$&all others\\
\hline
X & S & NS & S & NS\\
Y & NS & S & S & NS\\
\end{tabular}
\caption{The two states $X$ and $Y$ 
considered in the independent load condition.
The status of the contacts in the sets $\mathbb{X}$, $\mathbb{Y}$,
and $\mathbb{S}$ are given (S=sliding, NS=non-sliding).  
All contacts not in these sets are non-sliding
in both states.}
\label{XYTable}
\end{table}
Let us consider two different states $X$ and $Y$, each with a different
set of sliding contacts.  Let $\mathbb{S}$ be the set of contacts
sliding in both states, $\mathbb{X}$ be set of sliding contacts
unique to $X$ and $\mathbb{Y}$ be those unique to $Y$ 
(see Table~\ref{XYTable}).  Let $\mathbf{v}^X$
be the velocities in state $X$ and $\mathbf{v}^Y$ be those in $Y$.  Similarly,
$\mathbf{k}^X = \mathbf{k}(\mathbb{S}\cap\mathbb{X})$ and
$\mathbf{k}^Y = \mathbf{k}(\mathbb{S}\cap\mathbb{Y})$.  If
the externally applied force is independent of contact status,
\begin{equation}
\frac{d\mathbf{f}_\mathrm{ext}}{dt} = \mathbf{k}^X \mathbf{v}^X =
  \mathbf{k}^Y \mathbf{v}^Y,
\label{samesolution}
\end{equation}
or both $\mathbf{v}^X$ and $\mathbf{v}^Y$ are the velocities caused by
the same external forces, but with different stiffness matrices.

Eq.~(\ref{samesolution}) can be rewritten
\begin{equation}
\mathbf{k}^X\mathbf{v}^X-\mathbf{k}^Y\mathbf{v}^Y =
\mathbf{cK}\left[\mathbf{S}^X\mathbf{V}^X-\mathbf{S}^Y\mathbf{V}^Y\right] = 0.
\end{equation}
Now let us multiply this equation from the left by 
$(\mathbf{v}^X-\mathbf{v}^Y)^T$:
\begin{equation}
\left[\mathbf{V}^X-\mathbf{V}^Y\right]^T\mathbf{K}
\left[\mathbf{S}^X\mathbf{V}^X-\mathbf{S}^Y\mathbf{V}^Y\right] = 0.
\label{CommonSolution}
\end{equation}
This again is simply a sum over contacts:  
\begin{equation}
\sum_{\alpha=1}^M
\left[\underline{V}^X_\alpha-\underline{V}^Y_\alpha\right]^T\underline{K}_\alpha
\left[\underline{S}^X_\alpha\underline{V}^X_\alpha-\underline{S}^Y_\alpha\underline{V}^Y_\alpha\right] = 0.
\end{equation}
There will be four types of contributions, corresponding to the four
columns in Table~\ref{XYTable}:
\begin{itemize}
\item Contacts which slide in $X$ but not in $Y$ (the set $\mathbb{X}$),
\item Contacts which slide in $Y$ but not in $X$ (the set $\mathbb{Y}$),
\item Contacts which are sliding in both $X$ and $Y$ (the set $\mathbb{S}$),
\item Contacts which are non-sliding in both $X$ and $Y$.
\end{itemize}
For the last two classes of contacts, $\underline{S}^X=\underline{S}^Y$,
so their contributions here will be the same as to the quadratic form.  
For contacts $\alpha\in\mathbb{X}$ the contribution is:
\begin{equation}
K_n(V_{n,\alpha}^X-V_{n,\alpha}^y)^2 
  + (\mu K_n V_{n,\alpha}^X \sgn D_{t,\alpha} + K_t V^Y_{t,\alpha})(V_{t,\alpha}^X-V_{t,\alpha}^y).
\end{equation}
Defining 
$\hat F^{XY}_\alpha = \hat F^X_\alpha - \hat F^Y_\alpha$,
this quantity can be rewritten as
\begin{equation}
Q^\mathrm{(NS)}_\alpha(\mathbf{v}^X-\mathbf{v}^Y)
   - \hat F^{XY}_\alpha \tilde V^X_\alpha.
\end{equation}
In the same way, the contribution of contacts $\alpha\in\mathbb{Y}$ is
\begin{equation}
Q^\mathrm{(NS)}_\alpha (\mathbf{v}^X-\mathbf{v}^Y)
   + \hat F^{XY}_\alpha \tilde V^Y_\alpha.
\end{equation}
Thus Eq.~(\ref{CommonSolution}) becomes
\begin{equation}
Q(\mathbf{v}^X-\mathbf{v}^Y,\mathbb{S})  =
 \sum_{\alpha\in\mathbb{X}} \hat F^{XY}_\alpha \tilde V^X_\alpha
 - \sum_{\alpha\in\mathbb{Y}} \hat F^{XY}_\alpha \tilde V^Y_\alpha.
 \label{LoadIndpendent}
\end{equation}

\subsection{Small numbers of sliding contacts}
\label{smallMc}

\begin{table}
\begin{tabular}{c|ccccc}
&\multicolumn{2}{c}{Contact status}&
\multicolumn{2}{c}{$\sgn\tilde V$}&Consistency\\
State&$\quad\beta\quad$&$\quad\gamma\quad$&$\quad\beta\quad$&
$\quad\gamma\quad$&requirements\\
\hline
A&NS&NS&$1$&$\sigma_2$& $\beta$, $\gamma$ not critical\\
B&S&NS&$1$&$1$&$\gamma$ not critical\\
C&S&S&$\sigma_1$&$1$&$\sigma_1=1$\\
D&NS&S&$\sigma_1$&$\sigma_2$&$\sigma_1=-1$, $\sigma_2=1$\\
\end{tabular}
\caption{The states considered in Sec.~\ref{smallMc}.
Contacts $\beta$ and $\gamma$ are critical, so there are four possible
states labeled by $A$, $B$, $C$, and $D$.  The table gives the status
of each critical contact in each state (S=sliding, NS=non-sliding).
Also given are the signs of $\tilde V$.  $\tilde V^A_\beta$ 
and $\tilde V^B_\gamma$
are known to be positive, because it is assumed that $\beta$ becomes
critical while the packing is in state $A$, and $\gamma$ becomes critical
while in state $B$.  The other values are deduced from the one contact
status change theorem given at the end of Sec.~\ref{oneSlides}.  $\sigma_1,
\sigma_2=\pm1$ are unknown, but used here to show relations between
different states.}
\label{states}
\end{table}

In preparation for treating the general case, we will consider the
problem of a packing that may slide at two different contacts $\beta$
and $\gamma$.  The four different possible states are shown in 
Table~\ref{states}, and labeled $A$, $B$, $C$, and $D$.  We will use
superscripts to indicate quantities belonging to each state.  For example
$\mathbf{v}^A$ are the particle velocities in state $A$ and 
$\mathbf{S}^C$ is the status matrix in state $C$.

The system starts in state $A$ with no sliding
contacts.  Then contact $\beta$ reaches the boundary of the Coulomb
cone and becomes sliding, and the packing moves to state $B$.  Then
contact $\gamma$ reaches the boundary, the system moves to either
state $C$, where both $\beta$ and $\gamma$ slide, or to state $D$,
where only $\gamma$ slides.  

We will consider the questions of existence and
uniqueness.  For example, when $\beta$ becomes sliding, is it guaranteed
that $B$ is consistent?  If it were inconsistent, then contact $\beta$ should
become non-sliding.  But if it became non-sliding, the system would
return to state $A$, and $\beta$ would leave the Coulomb cone.
A solution would not exist.  The question of uniqueness
arises when contact $\gamma$ reaches the boundary.
If both states $C$ and $D$ were consistent,
the system could move to either $C$ or $D$, and the solution would
not be unique.  We will show that the consistent state exists and is unique.

\subsubsection{One sliding contact}
\label{oneSlides}

We first consider the transition from $A$ to $B$.  This transition
occurs when the contact $\beta$ reaches the boundary
of the Coulomb cone.  It starts somewhere within the cone, that is
with $\tilde F_\beta>0$ [see Eq.~(\ref{Coulombcondition})].  As the contact
moves toward the boundary, $\tilde F_\beta$ decreases and then vanishes
when $\beta$ reaches the boundary.  Therefore, Eq.~(\ref{FsVs}) requires
that
\begin{equation}
\tilde V^A_\beta > 0.
\label{betareach}
\end{equation}
Thus a $1$ is given in the top row of the third column of Table~\ref{states}.

One usually supposes without comment that the state $B$ is consistent.
But this is not obvious, because all particles change
their velocities when the state changes.  The state $B$ will be
consistent only if  $\tilde V^B_\beta>0$, and no one has shown 
that this must be so.

To show that $B$ is indeed consistent, let us apply the independent
load condition to the transition between
$A$ and $B$.  Setting $X=A$, $Y=B$, $\mathbb{S}=\mathbb{X}=\emptyset$,
and $\mathbb{Y}=\{\beta\}$, Eq.~(\ref{LoadIndpendent}) becomes
\begin{equation}
Q(\mathbf{v}^A-\mathbf{v}^B,\emptyset) = - \hat F^{AB}_\beta \tilde V^B_\beta.
\label{Qempty}
\end{equation}
If state $A$ is stable, the quadratic form must be positive, leading to
\begin{equation}
\hat F^{AB}_\beta \tilde V^B_\beta < 0,
\label{alpha_ineq}
\end{equation}
which gives us some information about the sign of $\tilde V^B_\beta$,
but also unfortunately involves the unknown quantity $\hat F^{AB}_\beta$.
More information can be obtained by requiring state $B$ to be stable:
\begin{equation}
Q(\mathbf{v}^A-\mathbf{v}^B,\{\beta\})> 0,
\label{Bstable}
\end{equation}
and after using Eq.~(\ref{stability2})
\begin{equation}
Q(\mathbf{v}^A-\mathbf{v}^B,\emptyset) -\hat F^{AB}_\beta
 ( \tilde V^A_\beta - \tilde V^B_\beta) > 0,
\label{Bstablebis}
\end{equation}
and finally using Eq.~(\ref{Qempty}):
\begin{equation}
\hat F^{AB}_\beta \tilde V^A_\beta < 0,
\label{beta_ineq}
\end{equation}
Together Eqs.~(\ref{beta_ineq}) and (\ref{alpha_ineq}) show that
$\tilde V^A_\beta$ and $\tilde V^B_\beta$ have the same sign.  
We already showed that $\tilde V^A_\beta>0$ in Eq.~(\ref{betareach}),
thus $\tilde V^B_\beta>0$ as well.  Therefore state $B$
is compatible, and the solution exists.

Before proceeding, let us pause to note that the reasoning we have just
employed does not depend on state $A$ being without sliding contacts.
Define $\mathbb{S}$ to be the set of contacts sliding in state
$A$.  If we simply replace the empty set in Eq.~(\ref{Qempty}) 
and (\ref{Bstablebis}) with 
$\mathbb{S}$, and $\{\beta\}$ with $\{\beta\}\cap\mathbb{S}$ in
Eq.~(\ref{Bstable}), the reasoning remains unchanged.  Thus we have
a general statement:
If two states $A$ and
$B$ differ only in the status of a single contact $\beta$,
then $\sgn \tilde V^A_\beta=\sgn \tilde V^B_\beta$.
This state implies that the consistency at contact $\beta$ must be
different in states $A$ and $B$.  If 
$\sgn\tilde V^A_\beta = \sgn\tilde V^B_\beta=-1$ (or if $\beta$
is not a critical contact) then
$A$ will be consistent at $\beta$ and $B$ will be inconsistent.  If
$\sgn\tilde V^A_\beta = \sgn\tilde V^B_\beta=-1$ then $A$
will be inconsistent and $B$ consistent.  Thus we have proved a
special case of the status change theorem:
\begin{quote}
\textsl{One contact status change theorem}: If two states $A$ and
$B$ differ only in the status of a single contact $\beta$,
then $\sgn \tilde V^A_\beta=\sgn \tilde V^B_\beta$, meaning
that they differ in the consistency at $\beta$.
\end{quote}
Using this theorem, we can now fill in the third and fourth columns
of Table~\ref{states}.  States $C$ and $D$ differ only in the status
of contact $\beta$, so $\sgn\tilde V^C_\beta=\sgn\tilde V^D_\beta=\sigma_1$.
States $A$ and $D$ differ only in the status of contact $\gamma$,
so $\sgn\tilde V^A_\gamma=\sgn\tilde V^D_\gamma=\sigma_2$.
For the same reason, $\sgn\tilde V^B_\gamma=\sgn\tilde V^C_\gamma$,
and in the next section, we show $\sgn\tilde V^B_\gamma=1$.

\subsubsection{Two sliding contacts}
\label{twoSlides}

Suppose now that the system is in state $B$, when a
second contact $\gamma$ reaches the boundary of the
Coulomb condition.  Following the same reasoning as at the
beginning of the preceding Sec.~\ref{oneSlides}, we see
\begin{equation}
\tilde V^A_\gamma > 0.
\end{equation}
Thus a $1$ is given in the second row of the fourth column of
Table~\ref{states}.  The one contact status change theorem tells
us we should immediately put a $1$ in the third row of the same
column.

Now let us consider the uniqueness of the globally consistent state.
When $\gamma$ becomes sliding,
the system can now move to either state $C$ where
both $\beta$ and $\gamma$ slide, or to $D$, where only $\gamma$
slides.  We know that the system cannot return to $A$,
because $\tilde V^A_\gamma>0$, so that state will no longer
be consistent.

The one contact status change theorem shows that
the solution is unique.  Only a single contact is
different between states $C$ and $D$, and the theorem states that
only one of these states can be consistent at $\beta$.
If $\sgn\tilde V^C_\beta = \sgn \tilde V^D_\beta = \sigma_1 =1$, then contact
$\beta$ must slide, and $C$ is consistent but not $D$.
On the other hand, if $\sigma_1 =-1$,
then contact $\beta$ must be non-sliding, and $D$ is consistent but not $C$.

Now let us check the existence of the solution.  Suppose
$\sigma_1=1$, meaning
the system must move to state $C$.  In order for this state
to be compatible, we must also have $\tilde V^C_\gamma>0$.  As shown
above, the one contact status change theorem guarantees this.

What happens if $\sigma_1=-1$?
In this case, the system must move to state $D$, but is this state
compatible?  The first order state theorem
says $\sgn\tilde V^D_\gamma=\sgn\tilde V^A_\gamma=\sigma_2$,
but this does not help prove compatibility, since $\sigma_2$
is unknown.  However, we do know that $\tilde V^B_\gamma>0$,
so let us now search for a way to relate $\tilde V^B_\gamma$ to
$\tilde V^D_\gamma$.  This cannot be done by the one contact
status change theorem, because $B$ and $D$ differ at two contacts.
Therefore, we must return to two hypotheses discussed in
Sec.~\ref{premise}, and deduce more information from them.

If we apply the independent load condition to
states $B$ and $D$, we obtain
\begin{equation}
Q(\mathbf{v}^B-\mathbf{v}^D,\emptyset) =
 \hat F^{BD}_\beta \tilde V^B_\beta - \hat F^{BD}_\gamma \tilde V^D_\gamma.
\end{equation}
Requiring all four states $A$, $B$, $C$, and $D$ to be stable leads to
the inequalities
\begin{eqnarray}
\hat F^{BD}_\beta \tilde V^B_\beta > \hat F^{BD}_\gamma \tilde V^D_\gamma,
	\label{stable1}\\
\hat F^{BD}_\beta \tilde V^D_\beta > \hat F^{BD}_\gamma \tilde V^D_\gamma,
	\label{stable2}\\
\hat F^{BD}_\beta \tilde V^B_\beta > \hat F^{BD}_\gamma \tilde V^B_\gamma,
	\label{stable3}\\
\hat F^{BD}_\beta \tilde V^D_\beta > \hat F^{BD}_\gamma \tilde V^B_\gamma.
	\label{stable4}
\end{eqnarray}
Now let us suppose that $\sgn\tilde V^B_\beta \ne \sgn\tilde V^D_\beta$
(i.e.~$\sigma_1=-1$).
This means that the left hand side of either Eq.~(\ref{stable1})
or (\ref{stable2}) must be negative.  The right hand side of these
two relations is identical, and must be less than some negative number:
\begin{equation}
F^{BD}_\gamma \tilde V^D_\gamma < 0.
\end{equation}
Similar reasoning with Eqs.~(\ref{stable3}) and (\ref{stable4}) leads to
\begin{equation}
F^{BD}_\gamma \tilde V^B_\gamma < 0.
\end{equation}
These two inequalities require that $V^D_\gamma$ and $V^B_\gamma$
have the same sign.  Thus we have shown
\begin{equation}
\sgn \tilde V^B_\beta \ne \sgn \tilde V^D_\beta
\quad \Rightarrow \quad
\sgn \tilde V^B_\gamma = \sgn \tilde V^D_\gamma.
\label{gammaeq}
\end{equation}
In the same way, one can begin with the assumption that
$V^D_\gamma$ and $V^B_\gamma$ have opposite signs, and obtain
\begin{equation}
\sgn \tilde V^B_\gamma \ne \sgn \tilde V^D_\gamma
\quad \Rightarrow \quad
\sgn \tilde V^B_\beta = \sgn \tilde V^D_\beta.
\label{betaeq}
\end{equation}

Now let us use these results to determine the consistency of state $D$.
Using the information given in Table~\ref{states}, Eq.~(\ref{gammaeq})
becomes
\begin{equation}
\sigma_1 \ne 1 \quad \Rightarrow \quad \sigma_2 = 1.
\end{equation}
Recall that $\sigma_1=-1$ means that state $C$ is inconsistent,
and $\sigma_1=-1$, $\sigma_2=1$ is the condition required for
$D$ to be consistent.  Thus the consistent state always exists.

But Eqs.~(\ref{gammaeq}) and (\ref{betaeq}) also have a much more profound
implication.  When moving from state $B$ to state $D$, the sign of $\tilde V$
at both $\beta$ and $\gamma$ cannot change.  Note that this reasoning
holds even if there are other sliding contacts, as long as they remain
sliding in all four states $A$, $B$, $C$, and $D$.  Thus we have
a second special case of the status change theorem:
\begin{quote}
\textsl{Two contact status change theorem:} If two states $A$ and
$B$ differ only in the status of two contacts $\beta$ and $\gamma$,
then $\sgn \tilde V^A_\beta=\sgn \tilde V^B_\beta$ or
$\sgn\tilde V^A_\gamma=\sgn\tilde V^B_\gamma$, meaning that the
consistency at $\beta$ or $\gamma$ (or both) must change.
\end{quote}

\subsection{Many sliding contacts}
\label{bigMc}

\begin{table}
\begin{tabular}{c|cccccc}
State&$\;\mathbb{A}\backslash\mathbb{A}'\;$&$\quad\mathbb{A}'\quad$&
$\;\mathbb{B}\backslash\mathbb{B}'\;$&$\quad\mathbb{B}'\quad$&
$\quad\mathbb{S}\quad$& all others\\
\hline
A&S&S&NS&NS&S&NS\\
B&NS&NS&S&S&S&NS\\
C&NS&S&NS&S&S&NS
\end{tabular}
\caption{States for the proof of the status change theorem.  Two states
$A$ and $B$ are considered.  For the proof, a third state $C$ is
constructed.  Contacts in $\mathbb{A}$ are sliding in $A$, but not in
$B$, and contacts in $\mathbb{B}$ are sliding in $B$, but not in $A$.
Sets $\mathbb{A}'\subset\mathbb{A}$ and $\mathbb{B}'\subset\mathbb{B}$
are sliding in state $C$.  (Here S=sliding, NS=non-sliding).}
\label{StatChangeTable}

\end{table}

In this section, we prove the status change theorem.  
Let us restate it in this way:
\begin{quote}
\textsl{Status change theorem:} If two states $A$ and
$B$ differ only in the status of $n$ contacts $\beta_1, \ldots \beta_n$,
then for at least one contact $\beta_i$, we have
$\sgn \tilde V^A_{\beta_i}=\sgn \tilde V^B_{\beta_i}$, implying that
the consistency at $\beta_i$ must also change.
\end{quote}
Recall that in Sec.~\ref{ProofPreview}, we showed that this statement
is sufficient to prove that the globally consistent state exists and
is unique.  We now establish this theorem.

Consider two different states $A$ and $B$.  The contacts in
$\mathbb{A}$ are sliding in $A$ but not in $B$, and the contacts
in $\mathbb{B}$ are sliding in $B$ but not in $A$.  Let $\mathbb{S}$
contain contacts that are sliding in both states.  We want to show
that 
\begin{equation}
\sgn \tilde V^A_\alpha = \sgn \tilde V^B_\alpha \mbox{ for at least one }
\alpha\in\mathbb{A}\cap\mathbb{B}.
\label{CorrectHypothesis}
\end{equation}
We will begin by assuming the contrary:
\begin{equation}
\sgn \tilde V^A_\alpha \ne \sgn \tilde V^B_\alpha \mbox{ for all }
\alpha\in\mathbb{A}\cap\mathbb{B},
\label{WrongHypothesis}
\end{equation}
and show that this leads to a contradiction.

The independent load condition Eq.~(\ref{LoadIndpendent}) implies
\begin{equation}
Q(\mathbf{v}^A-\mathbf{v}^B,\mathbb{S})  =
 \sum_{\alpha\in\mathbb{A}} \hat F^{AB}_\alpha \tilde V^A_\alpha
 - \sum_{\alpha\in\mathbb{B}} \hat F^{AB}_\alpha \tilde V^B_\alpha.
 \label{IndLoad1}
\end{equation}

Now let us assume stability for a third state $C$.  The following
contacts shall be sliding in $C$:
\begin{itemize}
\item All contacts in $\mathbb{S}$, who are sliding in both states
$A$ and $B$.
\item Some contacts that are sliding in $A$ but not in $B$.  Let
$\mathbb{A}'\subset\mathbb{A}$ denote these contacts.
\item Some contacts that are sliding in $B$ but not in $A$.  Let
$\mathbb{B}'\subset\mathbb{B}$ denote these contacts.
\end{itemize}
Table~\ref{StatChangeTable} summarizes this information.

Let $\mathbb{C}=\mathbb{A}\cup\mathbb{B}$ and 
$\mathbb{C}'=\mathbb{A}'\cup\mathbb{B}'$.
Note that $\mathbb{C}'$ can be any subset of $\mathbb{C}$.
The stability condition for state $C$ is
\begin{eqnarray}
Q(\mathbf{v}^A-\mathbf{v}^B,\mathbb{S}\cup\mathbb{C}') &>& 0,\cr 
Q(\mathbf{v}^A-\mathbf{v}^B,\mathbb{S})
+\sum_{\alpha\in\mathbb{C}'} 
            \hat F^{AB}_\alpha(\tilde V^B_\alpha-\tilde V^A_\alpha)&>& 0,
\end{eqnarray}
Combining this with Eq.~(\ref{IndLoad1}) yields 
\begin{eqnarray}
\sum_{\alpha\in\mathbb{A}'} \hat F^{AB}_\alpha \tilde V^B_\alpha +
\sum_{\alpha\in\mathbb{A}\backslash\mathbb{A}'} \hat F^{AB}_\alpha \tilde V^A_\alpha&
&\cr
-\sum_{\alpha\in\mathbb{B}'} \hat F^{AB}_\alpha \tilde V^A_\alpha -
\sum_{\alpha\in\mathbb{B}\backslash\mathbb{B}'} \hat F^{AB}_\alpha \tilde V^B_\alpha &> 0
\label{stabilities}
\end{eqnarray}
Note that $\mathbb{A}'$ and $\mathbb{B}'$ are arbitrary, so there is a large
number of such relations.

To write these relations in a more compact form, we define
\begin{equation}
\phi^A_\alpha = \left\{ \begin{array}{ll}
F^{AB}\tilde V^A_\alpha,&\mbox{for } \alpha\in\mathbb{A},\\
-F^{AB}\tilde V^A_\alpha,& \mbox{for } \alpha\in\mathbb{B},
\end{array}\right.
\end{equation}
with an analogous definition for $\phi^B_\alpha$.  
Now the relations Eq.~(\ref{stabilities}) can be written
\begin{equation}
\sum_{\alpha\in\mathbb{C}'} \phi^B_\alpha + 
\sum_{\alpha\in\mathbb{C}\backslash\mathbb{C}'} \phi^A_\alpha > 0.
\label{stabilitiesbis}
\end{equation}
And the hypothesis Eq.~(\ref{WrongHypothesis}) becomes
\begin{equation}
\sgn \phi^A_\alpha \ne \sgn \phi^B_\alpha \mbox{ for all }
\alpha\in\mathbb{C},
\label{WrongHypothesisbis}
\end{equation}
To show that Eq.~(\ref{WrongHypothesisbis}) leads to a contradiction, 
it suffices to show
\begin{equation}
\sum_{\alpha\in\mathbb{C}'_i} \phi^B_\alpha + 
\sum_{\alpha\in\mathbb{C}_i\backslash\mathbb{C}'_i} \phi^A_\alpha > 0
\mbox{ for }1\le i \le n.
\label{iinequality}
\end{equation}
where $n$ is the number of elements in $\mathbb{C}$, and 
$\mathbb{C}_i$ is a subset of $\mathbb{C}$
that contains exactly $i$ elements, and $\mathbb{C}_i'$ is
any subset of $\mathbb{C}_i$. 

The case $i=1$ of Eq.~(\ref{iinequality}) 
contradicts the hypothesis Eq.~(\ref{WrongHypothesisbis}).
To see this, let $\mathbb{C}_1=\{ \alpha \}$.  Choosing 
$\mathbb{C}_1'=\emptyset$ in Eq.~(\ref{iinequality}) leads
to $\phi_\alpha^A>0$, and choosing $\mathbb{C}'_1= \{ \alpha \}$ leads
to $\phi_\alpha^B>0$.  Thus $\phi_\alpha^A$ and $\phi_\alpha^B$ do
not have opposite signs, as assumed in Eq.~(\ref{WrongHypothesisbis}).
This means that Eq.~(\ref{WrongHypothesis}) is false, and 
Eq.~(\ref{CorrectHypothesis}) must be true.  Eq.~(\ref{CorrectHypothesis})
is equivalent to the status change theorem.

We now show Eq.~(\ref{iinequality}) by induction, beginning with
$i=n$ and proceeding to $i=1$.
The case $i=n$ is trivial, since $\mathbb{C}_n = \mathbb{C}$.
In this case, Eqs.~(\ref{stabilitiesbis}) and (\ref{iinequality})
identical.

Now let show that if Eq.~(\ref{iinequality}) holds for $i+1$,
then it holds for $i$ also. 
Choose a contact $\beta$ such that $\beta\not\in\mathbb{C}_i$, but
$\beta\in\mathbb{C}$.  By the hypothesis, Eq.~(\ref{iinequality})
holds for $\mathbb{C}_{i+1}=\mathbb{C}_i\cup\{\beta\}$.
Next, we make two different choices for $\mathbb{C}_{i+1}'$ and
apply Eq.~(\ref{iinequality}).
First we choose $\mathbb{C}_{i+1}'=\mathbb{C}_i'$ and obtain
\begin{equation}
\sum_{\alpha\in\mathbb{C}_i'} \phi^B_\alpha +
\sum_{\alpha\in\mathbb{C}_i\backslash\mathbb{C}_i'} \phi^A_\alpha
  > -\phi^A_\beta,
\end{equation}
and next we choose $\mathbb{C}_{i+1}'=\mathbb{C}_i'\cup\{\beta\}$:
\begin{equation}
\sum_{\alpha\in\mathbb{C}_i'} \phi^B_\alpha +
\sum_{\alpha\in\mathbb{C}_i\backslash\mathbb{C}_i'} \phi^A_\alpha
  > -\phi^B_\beta,
\end{equation}
Note the parallel between these conditions and 
Eqs.~(\ref{stable1}) through (\ref{stable4}).
By Eq.~(\ref{WrongHypothesisbis}) either $\phi^A_\beta$ or
$\phi^B_\beta$ is negative.  Therefore, the only way for both
of these inequalities to hold is if the sums are positive:
\begin{eqnarray}
\sum_{\alpha\in\mathbb{C}_i'} \phi^B_\alpha +
\sum_{\alpha\in\mathbb{C}_i\backslash\mathbb{C}_i'} \phi^A_\alpha >0
\end{eqnarray}
This completes the induction step, and thus the proof.

\section{Discussion and Conclusion}
\label{conclusion}

\subsection{How restrictive are the assumptions?}

This work considered only circular particles.  The conclusions are
probably not modified if other shapes are considered.  The particle
shape most strongly affects the geometric stiffness, which is
neglected because of the quasi-rigid assumption.  The moment of
inertia plays only a small role, because it is eliminated in the
quasi-static approximation.  To accommodate particles of different
shapes, the particle radius $r$ in Eq.~(\ref{feqn}) must depend
on the contact, and the torque may also depend on the normal
force.  This requires modifying the matrix $\mathbf{c}$.  But
none of these should alter our considerations of mechanism indeterminacy,
nor alter the two premises of the proof used to resolve
contact status indeterminacy.

Another assumption that seems quite restrictive is the use of the
linear force law in deriving the stiffness matrix.  
This is no restriction, because this paper revolves around the
question of what occurs at one point in time, when the system
must adjust the status of the contacts.  Therefore, one could
always linearize the force law around the positions of the
particles.

A second assumption is that all possible states must be stable. 
This assumption is reasonable because if there is an unstable state,
the packing will probably collapse, rendering the question of uniqueness
irrelevant.  The state that is most likely to be unstable is the one where
all $M_c$ critical contacts are sliding.  
This is so, because the contributions of sliding contacts to the
quadratic form are on the average negative \cite{PG}.  Furthermore,
the only way to obtain
instability is for negative contributions of the sliding contacts
to the quadratic form outweigh the positive contributions of
the non-sliding contacts [see discussion of Eq.~(\ref{SlidingNegative})].
Now, the packing is always close to this state, because when a contact becomes
non-sliding, it leaves the boundary of the Coulomb cone and becomes
non-critical.  Therefore, when the assumption of stability for all
possible states is violated, we expect the packing to yield.

Furthermore, if there are vibrations in the system, they will be governed
by Eq.~(\ref{oscillate}).  These vibrations will cause the relative
motion at each contact to fluctuate.  At critical contacts, the
contact status will therefore switch between non-sliding and sliding.
Thus the packing will sample many different possible states, and if
it finds an unstable one, it may collapse.  Thus all states must
be stable in order to guarantee that the vibrations 
will damp out, meaning that this
is a necessary condition to obtain quasi-static balance.

Finally, let us remark that theory presented in Sec.~\ref{QQD} requires
some modifications to deal with opening and closing contacts.  To account
for a contact that opens, it is necessary to introduce the status ``open'',
and allow $D_n$ to become positive.  Another problem is presented by
contacts that are initially ``open'' but may later close.  In the theory, the
particle deformations are assumed to be infinitely small, so that
no two particles separated by a finite distance will ever touch.
This may lead to the omission of important effects when
the particle separations are very small, such as in a regular packing of
almost monodisperse spheres.

\subsection{Implications of the result}

This work suggests that the stiffness matrix, together with the 
principle of minimum sliding form a complete description of
quasi-static granular material.  Since the globally consistent state
always exists and is unique, there is no need to appeal to other
processes that have been left out of the model to decide between
various possible states.  Instead of rather brutally setting the
particle displacements to zero, as is done in various ``stress-only''
approaches to granular matter, one should consider
taking the ``quasi-rigid'' limit, where the stiffness of the particles
diverges, and the displacements become infinitesimally small, but
are not set to zero.  Taking this limit leads to the stiffness
matrix approach discussed in this paper.  This work supports
the conjecture that the quasi-rigid limit preserves all the necessary
physics needed to describe the quasi-static behavior.

Of course, there remain some open questions.  For example, the question
of opening and closing contacts has not been dealt with.  When a contact
reaches the apex of the Coulomb cone, it can then go into four different
states.  It can become non-sliding, and move into the interior of the
Coulomb cone.  Then, there are two distinct sliding states - each one 
corresponding to a different side of the Coulomb cone.  Finally, the 
contact can open.
Are these four states mutually exclusive, as we have shown to be
the case for the sliding and non-sliding states in this paper?

This work should also encourage the use of numerical methods based
on the stiffness matrix.  The problem faced by these methods is,
of course, to find the globally consistent contact status.  This work
shows that such a state always exists, and is unique.  Thus any way
to find the state is acceptable.  Furthermore, perhaps it is possible
to use the results of this paper to design intelligent strategies
for finding the globally consistent state.

\acknowledgments

We thank Jean-No\"el Roux for drawing our attention to 
contact status indeterminacy.  We also thank Matthew Kuhn and Katalin
Bagi for making their manuscripts available.  We acknowledge the support
of the European DIGA project HPRN-CT-2002-00220, and the Deutsche
Forschungsgemeinschaft through project GEP-HE-2732/8-1.

\bibliographystyle{prsty}

\end{document}